\documentstyle[prb,aps,epsf,psfig]{revtex}

\begin{document}
\twocolumn[\hsize\textwidth\columnwidth\hsize
           \csname @twocolumnfalse\endcsname
\title{Shifts of the nuclear resonance in the vortex
lattice in YBa$_2$Cu$_3$O$_7$}
\author{Pavel Lipavsk\'y and Jan Kol{\'a}{\v c}ek}
\address{Institute of Physics, Academy of Sciences,
Cukrovarnick\'a 10, 16258 Praha 6, Czech Republic}
\author{Klaus Morawetz}
\address{Max Planck Institute for the Physics of Complex
Systems, Noethnitzer Str. 38, 01187 Dresden, Germany}
\author{Ernst Helmut Brandt}
\address{Max-Planck-Institut f\"ur Metallforschung,
         D-70506 Stuttgart, Germany}
\maketitle
\begin{abstract}
The NMR and NQR spectra of $^{63}$Cu in the CuO$_2$
plane of YBa$_2$Cu$_3$O$_7$ in the superconducting state
are discussed in terms of the phenomenological theory of
Ginzburg-Landau type extended to lower temperatures.
We show that the observed spectra,
Kumagai {\em et al.}, PRB {\bf 63}, 144502 (2001),
can be explained by a standard theory of the Bernoulli
potential with the charge transfer between CuO$_2$ planes
and CuO chains assumed.
\end{abstract}
\vskip2pc]

\section{Introduction}
Already in 1969 Rickayzen has derived the thermodynamic
correction to the electrostatic Bernoulli potential
in superconductors and has proposed to use this correction
to study the pairing interaction.\cite{R69} Capacitive
measurements of the Bernoulli potential have not shown
any signal of this correction,
however.\cite{BK68,BM68,MB71,CS86,CS96} It turned out
that the thermodynamic correction is cancelled
by the surface dipole,\cite{LKMM01} therefore, it can
be accessed only by measurements probing the bulk
of the superconductor. For a long time, there was no
experimental method to fulfill this task so that the
thermodynamic correction to the electrostatic potential
and the related charge transfer have remained a theoretical
topic\cite{KK92,KF95,BFGLO96,KL01,LKMB02} with no
experimental justification.

The experiment aimed at the charge transfer in the bulk
of the superconductor has been performed only recently.
Kumagai, Nozaki and Matsuda \cite{KNM01} (KNM) reported
very precise measurements of the nuclear magnetic resonance
(NMR) and the nuclear quadrupole resonance (NQR) spectra
for two high-$T_c$ materials, a slightly overdoped
YBa$_2$Cu$_3$O$_7$ and an underdoped YBa$_2$Cu$_4$O$_8$.
Comparing NMR and NQR frequencies they found that the
magnetic field, present in NMR while absent in NQR,
causes changes of the electric field gradient (EFG) in
the vicinity of the nucleus. In this way they
first observed the charge transfer in the Abrikosov vortex
lattice. They also pointed out that the conventional
theory of the vortex charge completely fails to explain
the experimental data giving a wrong sign and too small
amplitude of the magnetic field effect on the EFG.

In this paper we discuss the KNM experiment in terms of
a phenomenological theory of Ginzburg-Landau (GL) type
proposed in Refs.~\onlinecite{KL01,LKMB02}. This theory
is extended by features necessary to describe a layered
structure of YBCO, in particular the CuO chains which
serve as charge reservoirs. Unlike in a conventional
superconductor, where the charge is transfered on a long
distance from the vortex core to the region between the cores,
in the YBCO the dominant role plays a local charge transfer
between the CuO$_2$ planes and the CuO chains. This local
transfer will be shown to explain the observed sign and
amplitude of the magnetic field effect on the EFG. At low
temperatures,
there are two important corrections, the space variation
of the magnetic field and the small but finite disorientation
of the axis $c$ in individual grains of the measured sample.

\subsection{Electric field gradient effect in the vortex
lattice}
Let us briefly recall the experiment of Kumagai, Nozaki and
Matsuda.\cite{KNM01} The $^{63}$Cu atom in the CuO$_2$ plane
subjected to the magnetic field, $B \parallel c$, with
a number of holes, $N$, has NMR frequencies
$\nu_1=\gamma_{\rm Cu}B$ and
\begin{equation} 
\nu_{2,3}=\gamma_{\rm Cu}B\mp\left(C+AN\right).
\label{NMR1}
\end{equation}
The first term describes the coupling of the nuclear
magnetic momentum to the magnetic field, $B=9.4$~T in
Ref.~\onlinecite{KNM01}. The nucleus $^{63}$Cu has spin
3/2 and its Zeeman coefficient is
$\gamma_{\rm Cu}= 11.312$~MHz/T.

The second term of (\ref{NMR1}) is the EFG effect. The
nucleus $^{63}$Cu has a cigar shape elongated in the
direction of its magnetic momentum and this electric
quadrupole feels gradients of the electrostatic field
due to a non-spheric charge of electrons and ions in its
vicinity. The largest gradient of the electric field
(the principal axis of the EFG tensor) is parallel to
the axis $c$. The density dependent contribution to
the EFG is attributed to the local interaction between
nucleus with $d$-states on the Cu site. The experimental
fit\cite{Y95,KNM01} to the EFG effect yields $C=25$~MHz
and $A=34$~MHz per hole.

The NQR frequency includes the asymmetry correction,
$\nu_{\rm NQR}=\nu_0\sqrt{1+\eta^2/3}$, where $\eta$ is
the difference between gradients in the $a$ and $b$
directions scaled to the gradient
in the $c$ direction. For the Cu atom in the plane of
YBa$_2$Cu$_3$O$_7$ the $a$-$b$ asymmetry is rather weak,
$\sqrt{1+\eta^2/3}\approx 1+2.6\times 10^{-4}$.
The frequency $\nu_0$ corresponds to the NMR in the zero
magnetic field,
$\nu_0=C+AN_0$,
where $N_0$ is the number of holes in the superconducting
non-magnetic state.

A combination of the resonance frequencies,
\begin{equation} 
\Delta\nu=\nu_0-
{1\over 2}\left(\bar\nu_3-\bar\nu_2\right),
\label{NMR2}
\end{equation}
would be zero if the magnetic field has no effect on the
number of holes on the Cu sites. Actually, this is the case
for temperatures above $T_c$. Below the critical temperature,
the shift $\Delta\nu$ is non-zero indicating changes of
the quadrupole coupling due to the effect of the magnetic
field on the hole density. Briefly, $\Delta\nu$ serves as
a sensor of the charge transfer in the Abrikosov vortex
lattice.\cite{KNM01}

\subsection{Charge transfer between chains and planes}
In this paper we evaluate the shift $\Delta\nu$. To
motivate our approach we want to recall the way in which
the experimental data are obtained. The measurement yields
distributions $F_{2,3}(\nu)$ of the NMR frequencies, see
Fig.~3 in Ref.~\onlinecite{KNM01}. The frequencies
$\bar\nu_{3,2}$, used by KNM in formula (\ref{NMR2}) are
positions of the maxima of $F_{2,3}(\nu)$. Since NMR lines
have a finite width $\Gamma$, each site of a local
frequency $\nu_{2,3}({\bf r})$ contributes to the
spectrum by the Lorentzian
\begin{equation} 
F_i(\nu)=\left\langle
{\Gamma\over (\nu-\nu_i)^2+
\Gamma^2}\right\rangle
\equiv
{1\over\Omega}\int\! d{\bf r}
{\Gamma\over (\nu-\nu_i({\bf r}))^2+
\Gamma^2}.
\label{NMR3}
\end{equation}
The integral runs over the 2D volume, $\Omega=\int d{\bf r}$,
covering the elementary cell of the Abrikosov vortex lattice. The
magnetic field $B$ and the density $N$ are functions of
${\bf r}$, the frequencies $\nu_{2,3}$ depend on ${\bf r}$
via (\ref{NMR1}). The experimental magnetic field is the mean
value, $B=\langle B\rangle$.

The finite line width is crucial for the interpretation of the
KNM experiment. The observed contributions of
the quadrupole shifts are less than 25~kHz which is much
smaller than the width of the spectral line $\Gamma\approx
140$~kHz.  In this limit, the maximum of the spectral line
is given by the mean value and the lowest statistical variations,
\begin{equation} 
\bar\nu_i=\langle\nu_i\rangle-{2\left\langle(\nu_i-
\langle\nu_i\rangle)^3\right\rangle\over\Gamma^2-6\left
\langle(\nu_i-\langle\nu_i\rangle)^2\right\rangle}.
\label{NMR5}
\end{equation}

The variations are important at low temperatures due to a strong
compression of the magnetic field in vortex cores resulting in
a strong space-dependence of the Zeeman effect. Close to the
critical temperature the variations vanish and the maxima of
the NMR lines approach the mean frequencies,
$\bar\nu_{2,3}\approx\langle\nu_{2,3}\rangle$. Close to $T_c$
one finds from (\ref{NMR1}) and (\ref{NMR2}) that
\begin{equation} 
\Delta\nu\approx A\left(N_0-\langle N\rangle\right).
\label{NMR6}
\end{equation}
According to (\ref{NMR6}), the shift $\Delta\nu$ reflects
how many holes have been removed from the CuO$_2$ plane. The
in-plane charge transfer contributes only via variations
giving, as shown below, a negligible correction at all temperatures.

\subsection{Plan of the paper}
The physical picture we assume here is based on properties
of the non-magnetic state of YBCO. The superconducting state
appears mainly in CuO$_2$ planes, in chains the gap is
induced by the proximity effect.
The transition to the superconducting state shifts the
chemical potential in the same manner as in conventional
superconductors.\cite{BFGLO96,LKMB02} This shift is large
in planes, where the pairing mechanism is located, while
it is negligible in chains. In equilibrium the electrochemical 
potential is constant so the unequal shift of the chemical 
potential results in the charge transfer between plains and
chains, until the equilibrium condition is established by the 
electrostatic potential.

The magnetic field partially suppresses the
superconducting component what leads to a partial backward 
charge transfer. 
In the numerical treatment we assume both mechanisms, the
transfer between chains and planes and the in-plane transfer.
We show that this backward transfer between chains and planes 
brings the dominant contribution to the shift $\Delta\nu$.

In the next section we introduce the Lawrence-Doniach model
of the layered superconductor. In Sec.~III we derive a set
of equations for the GL function, the vector potential, and
the electrostatic potential. The charge density is obtained
from the Poisson equation adapted to the layered structure.
In the Sec.~IV we evaluate the distributions $F_{2,3}(\nu)$ from the
frequencies $\nu_{2,3}({\bf r})$ given by (\ref{NMR1}) with
the theoretical values of $B({\bf r})$ and $N({\bf r})$ obtained
from the extended GL theory \cite{LKMB02}. In Sec.~V we
present numerical results and discuss the individual
contributions to the experimental data. We show that
the approximation (\ref{NMR6}) applies close to the critical
temperature, but at low temperatures the space variation of
the magnetic field leads to appreciable corrections. In 
Subsec.~C
we compare our results with the model for conventional (not 
layered) superconductors and explain why the conventional model 
results in the reversed sign and too small amplitude of the 
line shift. In Subsec.~D we include a correction for samples 
made of $c$-oriented grains and show that the deviations from 
the ideal orientation are important at low temperatures. 
Sec.~VI contains a summary.

\section{Lawrence-Doniach model}
The model of layered superconductors has been proposed by
Lawrence and Doniach\cite{LD71}, see also Ref.~\onlinecite{B90}.
We associate the direction $z$ with
the $c$ axis of the YBCO and introduce a layer index $j$ and
an additional index $\pm$ which specifies the CuO$_2$ layer
position, $z_j^\pm=jD\pm {1\over 2}D_{pp}$,
where $D=11.65$~\AA\ is the period along the axis $c$ and
$D_{pp}=3.17$~\AA\ is the distance between two neighboring
planes. Positions of layers of chains are
$z_j^c=jD+{1\over 2}D.$

In the spirit of the tight-binding description, the variable $z$ is
replaced by the layer index. The wave function becomes a function
of the layer index, $\psi(x,y,z)\to\psi_j^{\pm,c}({\bf r})$,
and the 2D coordinate ${\bf r}\equiv (x,y)$.

\subsection{Kinetic energy}
The kinetic energy is composed of parts parallel to the planes and
the Josephson coupling along the $c$ axis. The kinetic part of
the free energy for the motion in the layer is treated within
the isotropic effective mass,
\begin{equation}
f_{\rm kin}^{j\iota}={1\over 2m_\iota^*}
\left|\left(-i\hbar\nabla-e^*{\bf A}_j^\iota\right)
\psi_j^\iota\right|^2,
\label{C1}
\end{equation}
where ${\bf A}_j^\iota$ is the in-plane component of the 3D
vector potential ${\bf A}(x,y,z_j^\iota)$. The masses in both
planes are identical, $m^*_\pm=m^*_p$. For simplicity we
assume the 2D effective mass $m^*_c$ in the chain layer to be
isotropic, too.

The motion between layers is represented by the Josephson
coupling,
\begin{eqnarray}
f_{\rm kin}^{j,+-}&=&J_{pp}\left|
\psi_j^+-{\rm e}^{-i\theta_j^{+-}}\psi_j^-\right|^2,
\nonumber\\
f_{\rm kin}^{j,c+}&=&J_{cp}\left|
\psi_j^c-{\rm e}^{-i\theta_j^{c+}}\psi_j^+\right|^2,
\nonumber\\
f_{\rm kin}^{j,-c}&=&J_{cp}\left|
\psi_j^--{\rm e}^{-i\theta_j^{-c}}\psi_{j-1}^c\right|^2.
\label{C2}
\end{eqnarray}
The phase factors $\theta$ of the off-diagonal terms are given by the
$c$-component of the 3D vector potential. For ${\bf B}
\parallel c$ we can use a gauge in which the $c$ component
of the vector potential is zero. In this gauge, all
three phases are zero, $\theta_j^{\iota\iota'}=0$. The total
kinetic free energy is a sum over all layers,
${\cal F}_{\rm kin}=\int d{\bf r}\sum_{j,\iota}
(f_{\rm kin}^{j\iota}+f_{\rm kin}^{j,\iota\iota'})$.

\subsection{Condensation energy}
Following Bardeen\cite{B54,B55} we use the Gorter-Casimir
free energy to describe the condensation. We assume that
the energy released by the formation of Cooper pairs is
located in CuO$_2$ planes. In terms of the wave function the
contribution of planes to the free energy reads\cite{LKMB02}
\begin{equation}
f_{\rm con}^{j\pm}=-\varepsilon_{\rm con}\varpi_j^\pm-
{1\over 2}\gamma_pT^2\sqrt{1-\varpi_j^\pm},
\label{con1}
\end{equation}
where $\gamma_p$ is the plane contribution to Sommerfeld's
$\gamma$ (linear coefficient of the electronic specific heat),
and
\begin{equation}
\varpi_j^\pm={2|\psi_j^\pm|^2\over
2|\psi_j^\pm|^2+n_{\rm nor}^{j,\pm}}
\label{con2}
\end{equation}
is the fraction of superconducting holes in the given plane.
The denominator of (\ref{con2}) is the total density of
pairable holes, $n_j^\iota=2|\psi_j^\iota|^2+
n_{\rm nor}^{j,\iota}$, which is a sum of the superconducting density
and the density of normal holes.

In the chains we treat the condensation energy as negligible.
The free energy thus includes only the contribution due to
the reduced entropy,
$f_{\rm con}^{jc}=-{1\over 2}\gamma_cT^2\sqrt{1-\varpi_j^c}$.
The meaning of symbols is analogous to the case of planes.
Again,  the sum over all layers reads ${\cal F}_{\rm con}=\int d{\bf r}\sum_{j\iota}
f_{\rm con}^{j\iota}$.

\subsection{Coulomb interaction}
The total density of holes is linked to the 2D density of
charge, $\rho_j^\iota=en_j^\iota+\rho_{\rm latt}^\iota$,
which creates the electrostatic potential,
\begin{equation}
\varphi_j^\iota({\bf r})={1\over 4\pi\epsilon}\int d{\bf r}'
\sum_{j'\iota'}{\rho_{j'}^{\iota'}({\bf r}')
\over\sqrt{|{\bf r'-r}|^2+(z_{j'}^{\iota'}-z_j^\iota)^2}}.
\label{C6}
\end{equation}
The Coulomb energy is
${\cal F}_C={1\over 2}\int d{\bf r}\sum_{j\iota}
\rho_j^\iota\varphi_j^\iota$.

It is advantageous to treat the electrostatic potential in
the momentum representation,
\begin{equation}
\varphi_\iota({\bf k})=\int d{\bf r}\,
\varphi_\iota({\bf r})
\,{\rm e}^{-i{\bf kr}}.
\label{C8}
\end{equation}
We have moved the superscript $\iota$ into subscript and
lifted the index $j$ since all equivalent layers are
at identical potentials, $\varphi_j^\iota=\varphi_m^\iota$.
The electrostatic potential then reads
\begin{equation}
\varphi_\iota({\bf k})={1\over\epsilon D}\sum_{\iota'}
L^{-1}_{\iota\iota'}\rho_{\iota'}({\bf k}),
\label{C9}
\end{equation}
with elements of the inverse Laplace operator
\begin{eqnarray}
L^{-1}_{++}=L^{-1}_{--}&=&L^{-1}_{cc}=
{D\over 2k}{1+{\rm e}^{-kD}\over 1-{\rm e}^{-kD}}
\nonumber\\
L^{-1}_{+-}=L^{-1}_{-+}&=&
{D\over 2k}{{\rm e}^{-kD_{pp}}+{\rm e}^{-k(D-D_{pp})}
\over 1-{\rm e}^{-kD}}
\nonumber\\
L^{-1}_{c\pm}=L^{-1}_{\pm c}&=&
{D\over 2k}{{\rm e}^{-kD_{cp}}+
{\rm e}^{-k(D_{cp}+D_{pp})}
\over 1-{\rm e}^{-kD}}.
\label{R1}
\end{eqnarray}
Here $D_{cp}={1\over 2}(D-D_{pp})$ denotes the distance
between adjacent chains and planes.

When inverted, the relation (\ref{C9}) yields the density of
charge in terms of the electrostatic potential,
\begin{equation}
\rho_\iota({\bf k})=\epsilon D\sum_{\iota'}L_{\iota\iota'}
\varphi_{\iota'}({\bf k}).
\label{R5}
\end{equation}
In our approach the electrostatic potential is evaluated
from the request of constant electrochemical potential, and
the relation (\ref{R5}) is used to evaluate the charge transfer.

\section{Equations of motion}
The total free energy is the sum of the components discussed
above and the Helmholtz free energy of the magnetic
field, ${\cal F}_M=\int d{\bf r}dz{1\over 2\mu_0}({\bf B}-
{\bf B}_0)^2$. Note that the magnetic energy does not have a
layered structure but it is given by a 3D integration.
The magnetic field is the 3D rotation of the vector potential,
${\bf B}={\rm rot}{\bf A}$. 

The total free energy, 
${\cal F}={\cal F}_{\rm kin}+{\cal F}_{\rm con}+
{\cal F}_C+{\cal F}_M$, 
is a functional of the wave function
$\psi$, the vector potential ${\bf A}$ and the normal density
$n_{\rm nor}$. By variations with respect to these functions
one arrives at equations of motion. For the conventional
superconductor, this procedure is described in
Ref.~\onlinecite{LKMB02}. We do not present its
straightforward modification for the layered structure but
discuss directly the results.

\subsection{Maxwell equation}
The functional variation $\delta{\cal F}/\delta{\bf A}$ yields
the Maxwell equation,
\begin{equation}
\nabla^2{\bf A}=-\mu_0\sum_{j\iota}
\delta(z-z_j^\iota)\,{\bf j}_j^\iota .
\label{C15}
\end{equation}
The right hand side includes the in-plane 2D currents,
\begin{equation}
{\bf j}_j^\iota={e^*\over m_\iota^*}{\rm Re}\,\bar\psi_j^\iota
\left(-i\hbar\nabla-e^*{\bf A}_j^\iota\right)\psi_j^\iota.
\label{C14}
\end{equation}
In principle, Josephson currents in the $c$ direction may be
also present. From
the translational and mirror symmetries it follows that
no Josephson currents flow between two neighboring CuO$_2$
planes. Assuming very small Josephson coupling between 
the chains and planes we neglect also the current 
between these layers.

The Maxwell equation (\ref{C15}) has the integral form
\begin{equation}
{\bf A}({\bf r},z)={\mu_0\over 4\pi}\int d{\bf r}'
\sum_{j'\iota'}{{\bf j}_{j'}^{\iota'}({\bf r}')
\over\sqrt{|{\bf r'-r}|^2+(z_{j'}^{\iota'}-z)^2}}.
\label{C16a}
\end{equation}
For $z=z_j^\iota$ the vector potential in the Fourier
representation simplifies to
\begin{equation}
{\bf A}_\iota({\bf k})={\mu_0\over D}\sum_{\iota'}
L^{-1}_{\iota\iota'}{\bf j}_{\iota'}({\bf k}).
\label{C16}
\end{equation}

For the YBCO with its large London penetration depth,
$\lambda_{\rm Lon}=14000$~\AA, the vector potential changes
only negligibly along the $c$ axis.
Indeed, the applied magnetic field 9.4~T results in a vortex
distance of 157~\AA, which is small compared to $\lambda_{\rm
Lon}$. The space modulation of the magnetic field is thus
covered by the lowest Fourier components of the Abrikosov
lattice. Since the vortex distance is large on the scale
of the YBCO lattice constant $D$, one can take the long
wave length limit, $k\to 0$, in which (\ref{C16}) reads
\begin{equation}
{\bf A}({\bf k})={\mu_0\over k^2}{\bf j}({\bf k}),~~~~~~~
{\bf j}({\bf k})={1\over D}
\sum_{\iota'}{\bf j}_{\iota'}({\bf k}).
\label{C16k0}
\end{equation}
In the approximation (\ref{C16k0}) the vector potential is
identical in chain and plane layers so that we can skip
its index.

\subsection{Schr\"odinger equation}
The Schr\"odinger equation of a system consisting of three
periodically repeated layers
has a complicated notation. We first write down this equation
in its general form and then reduce it by symmetries and the
approximation of weak Josephson coupling between chains and
planes. The result oriented reader can skip reading details
of the equations in this section.

The variation $\delta{\cal F}/\delta\bar\psi_j^\iota$ results
in the set of Lawrence-Doniach equations
\begin{eqnarray}
&&J_{pp}\left(\psi_j^+-\psi_j^-\right)+
J_{cp}\left(\psi_j^+-\psi_j^c\right)
\nonumber\\
&+&{1\over 2m^*_p}\left(-i\hbar\nabla-e^*{\bf A}
\right)^2\psi_j^++\chi_j^+\psi_j^+=0,
\nonumber\\
&&J_{pp}\left(\psi_j^--\psi_j^+\right)+
J_{cp}\left(\psi_j^--\psi_{j-1}^c\right)
\nonumber\\
&+&{1\over 2m^*_p}\left(-i\hbar\nabla-e^*{\bf A}
\right)^2\psi_j^-+\chi_j^-\psi_j^-=0,
\nonumber\\
&&J_{cp}\left(\psi_j^c-\psi_{j+1}^-\right)+
J_{cp}\left(\psi_j^c-\psi_j^+\right)
\nonumber\\
&+&{1\over 2m^*_c}\left(-i\hbar\nabla-e^*{\bf A}
\right)^2\psi_j^c+\chi_j^c\psi_j^c=0.
\label{C16LD}
\end{eqnarray}
We have used zero phases of the Josephson terms.

For ${\bf B}\parallel c$, all CuO$_2$ planes are identical,
therefore, $\psi_j^+=\psi_j^-=\psi_{j+1}^-\equiv\psi_p$.
The chain layers are also identical,
$\psi_j^c=\psi_{j-1}^c\equiv\psi_c$. This
simplifies the set (\ref{C16LD}) to two 2D Schr\"odinger
equations
\begin{eqnarray}
J_{cp}\left(\psi_p-\psi_c\right)
&+&{1\over 2m^*_p}\left(-i\hbar\nabla-e^*{\bf A}
\right)^2\psi_p+\chi_p\psi_p=0,
\nonumber\\
2J_{cp}\left(\psi_c-\psi_p\right)
&+&{1\over 2m^*_c}\left(-i\hbar\nabla-e^*{\bf A}
\right)^2\psi_c+\chi_c\psi_c=0.
\nonumber\\
\label{C16LDr}
\end{eqnarray}

The effective potentials,
\begin{eqnarray}
\chi_p&=&-2{\varepsilon_{\rm con}\over n_p}+
{\gamma_p T^2\over 2n_p}{1\over\sqrt{1-
{2|\psi_p|^2\over n_p}}},
\nonumber\\
\chi_c&=&{\gamma_c T^2\over 2n_c}{1\over\sqrt{1-
{2|\psi_c|^2\over n_c}}},
\label{chi1}
\end{eqnarray}
follow from the variation of the condensation energy.
The potential in chains consists of the entropy
term only.

If one neglects the effect of the charge transfer on the
material parameters $(n_p,n_c,\varepsilon_{\rm con},
\gamma_p,\gamma_c)$, the set of equations (\ref{C16k0}) and
(\ref{C16LDr}-\ref{chi1}) is closed. As in the ordinary
GL theory this set describes the magnetic properties of the
system. In fact, the transfered density of holes is very
small compared to the total densities $n_p$ and $n_c$. Therefore
in the first step, the magnetic field ${\bf B}$ and the wave
function $\psi$ can be evaluated from this set with no
regards to the electrostatic phenomena. The charge transfer
is evaluated in the second step from the wave function $\psi$.

\subsection{Electrostatic potential} From
variations of the free energy with respect to the
normal densities $n^{p,c}_{\rm nor}$ one finds the
electrostatic potential\cite{LKMB02}
\begin{eqnarray}
e\varphi_p&=&\chi_p{|\psi_p|^2\over n_p}+
{T^2\over 2}{\partial\gamma_p\over\partial n_p}
\sqrt{1-{2|\psi_p|^2\over n_p}}+
{\partial\varepsilon_{\rm con}\over\partial n_p}
{2|\psi_p|^2\over n_p}
\nonumber\\
e\varphi_c&=&\chi_c{|\psi_c|^2\over n_c}+
{T^2\over 2}{\partial\gamma_c\over\partial n_c}
\sqrt{1-{2|\psi_c|^2\over n_c}} \ .
\label{ephi}
\end{eqnarray}
The right hand sides represent local changes of the
chemical potential. Equations (\ref{ephi}) thus express
that the electrochemical potential remains constant.

One can see that the electrostatic potential in planes
differs from the potential in chains. At first glance,
these potentials differ by the term
\begin{equation}
{\partial\varepsilon_{\rm con}\over\partial n_p}\approx
{T_c^2\over 4}{\partial\gamma_p\over\partial n_p}+
{\gamma_pT_c\over 2}{\partial T_c\over\partial n_p}.
\label{ephi3}
\end{equation}
We have used the Gorter-Casimir formula for the
condensation energy, $\varepsilon_{\rm con}={1\over 4}
\gamma_pT_c^4$ and the assumption that the thermodynamic
properties of the superconducting phase are dominated by
planes.

For YBCO the term (\ref{ephi3}) is much larger
than any other term in the set (\ref{ephi}).
This is due to a rather strong dependence of the critical
temperature $T_c$ on the density of holes in planes.
Blatter's approximation\cite{BFGLO96} of the electrostatic
potential includes the second term of (\ref{ephi3}) only.
For simplicity we use Blatter's approximation by taking
\begin{equation}
\varphi_p={\gamma_pT_c\over e}{\partial T_c\over\partial
n_p}{|\psi_p|^2\over n_p},~~~~~~~~~~~~\varphi_c=0.
\label{ephi4}
\end{equation}
Using approximation (\ref{ephi4}) in the Poisson equation
(\ref{R5}) one obtains the 2D density of holes needed to
evaluate the electric field gradient effect on the NMR lines.

\subsection{Limit of weak Josephson coupling}
Within the approximation (\ref{ephi4}) the electrostatic potential
depends only on the wave function in planes. For a weak coupling
between chains and planes, $J_{cp}\to 0$, the set of two non-linear 
equations (\ref{C16LDr}) simplifies to a single equation,
\begin{equation}
{1\over 2m^*_p}\left(-i\hbar\nabla-e^*{\bf A}
\right)^2\psi+\chi_p\psi=0.
\label{SE}
\end{equation}
We have introduced a 3D wave function
$\psi=\sqrt{2\over D}\psi_p$,
in terms of which the current ${\bf j}$ has the usual GL form
\begin{equation}
{\bf j}={e^*\over m_p^*}{\rm Re}\,\bar\psi
\left(-i\hbar\nabla-e^*{\bf A}\right)\psi.
\label{C14psi}
\end{equation}
The wave function $\psi$ corresponds to the 3D density of
pairable holes, $n=2n_p/D$. As one expects, in the limit
of weak Josephson coupling, the magnetic properties are
described by the customary GL theory.

To summarize our approach, we use the numerical code developed
originally as a solver of the GL equations.\cite{B97} Its
modification to the Bardeen's approximation has been discussed
in Ref.~\onlinecite{LKMB02}. The output is the magnetic field
${\bf B}$ and the wave function $\psi$. The wave function is
scaled to its layered counterpart $\psi_p$ and used in (\ref{ephi4})
to provide us with the electrostatic potential in planes. From
the Poisson equation (\ref{R5}) we obtain the density
of holes in the CuO$_2$ planes as
\begin{equation} 
\rho_p({\bf k})={2k\epsilon\left(1+{\rm e}^{-kD}\right)\over
\left(1-{\rm e}^{-2kD_c}\right)
\left(1+{\rm e}^{-kD_{\rm int}}\right)}\varphi_p({\bf k}) .
\label{LS20}
\end{equation}

In our numerical treatment below we use the full expression
(\ref{LS20}). It should be noted, however, that the long
wave length limit $k\to 0$, i.e.,
$\rho_p({\bf k})={\epsilon\over D_c}\varphi_p({\bf k})$,
gives nearly the same result. As the relative errors due to
the long wave length approximation are of the order of
$10^{-3}$, one can say that the charge is transferred locally
from planes to chains. If the charge of planes and chains in
the vortex core is summed together, the vortex cores remain nearly
neutral.

\section{Electric field gradient effect}
Now we are ready to evaluate the NMR lines. The energy levels
of the nucleus with spin $I=3/2$ and spin component along
the magnetic field, $m=-3/2,-1/2,1/2,3/2$, read\cite{S78}
\begin{equation}
E_m=-\hbar\gamma_{\rm Cu}Bm+{e^2qQ\over 12}
\left(1-{3\over 2}\sin^2\theta\right)
\left(3m^2-{15\over 4}\right).
\label{Sl1}
\end{equation}
Here $Q$ is the quadrupole moment of the nucleus, $q$ is the
EFG which depends on the transfered charge, and $\theta$ is
the angle between the principal axis of the EFG tensor and
the magnetic field.

The NMR frequencies are differences between neighboring levels,
$\hbar\nu_1=E_{-1/2}-E_{1/2}$, $\hbar\nu_2=E_{1/2}-E_{3/2}$,
and $\hbar\nu_3=E_{-3/2}-E_{-1/2}$. The empirical formula
(\ref{NMR1}) corresponds to the magnetic field parallel to
the axis $c$. When the magnetic field declines from the axis
$c$ by angle $\theta$, the EFG effect on the NMR frequencies
is reduced,
\begin{equation}
\nu_{2,3}=\gamma_{\rm Cu}B\mp\left(C+AN\right)
\left(1-{3\over 2}\sin^2\theta\right).
\label{NMR1m}
\end{equation}

The amplitude of the local magnetic field changes in space,
$B=\langle B\rangle+\Delta B$. Although the deviations $\Delta B$
from the mean field are rather small, they bring an appreciable
contribution to the NMR frequencies on the scale of the shift
$\Delta\nu$.  The EFG effect depends on the angle $\theta$ given by
disorientations of grains in the sample. In the appendix we
show that the contribution of the magnetic field perpendicular
to the axis $c$ which appears even in a perfectly oriented
crystal due to the layered nature of diamagnetic currents is
negligible.

For a quantitative discussion of the EFG effect, it is
advantageous to compare the NMR frequency with its value in the
absence of diamagnetic currents,
$\nu_{2,3}=\nu^{\rm ref}_{2,3}+\Delta\nu_{2,3}$, where
\begin{equation}
\nu_{2,3}^{\rm ref}=\gamma_{\rm Cu}\langle B\rangle\mp AN_0.
\label{Sl3}
\end{equation}
The deviation caused by the diamagnetic currents and the
disorientation is
\begin{equation}
\Delta\nu_{2,3}=\gamma_{\rm Cu}\Delta B\mp
\left(A\Delta N-\nu_0{3\over 2}\sin^2\theta\right).
\label{Sl4}
\end{equation}

The density of holes per Cu site needed in (\ref{Sl4}) is
proportional to the charge density in the planes (\ref{LS20}),
\begin{equation} 
\Delta N({\bf r})={\Omega_{\rm Cu}\over e}\left(\rho_p
({\bf r})-\rho^0_p\right),
\label{LS8}
\end{equation}
where $\Omega_{\rm Cu}=14.88$~\AA$^2$ is the area per Cu
atom in a single plane. The density $\rho^0_p$ describes the
transfer of holes in the non-magnetic state. Since the
non-magnetic state is isotropic, i.e., it has only the
component $k=0$, from (\ref{ephi4}) and (\ref{LS20}) one finds
\begin{equation}
\rho^0_p={\epsilon\over D}
{|\psi_\infty|^2\over n_p^2}\gamma_p T_c^2
{\partial\ln T_c\over\partial\ln n_p},
\label{LS90}
\end{equation}
where $|\psi_\infty|^2={1\over 2}(1-t^4)n_p$ is the
non-magnetic value of the GL function in Bardeen's model,
$t=T/T_c$.

\subsection{Average over grain orientation}
So far we have discussed a single crystal. The sample
measured by KNM~\cite{KNM01} is made, however, from $c$
oriented grains. Of course,
there is a small but finite scatter in the orientation of
individual grains. The angle $\theta$ then varies from
grain to grain keeping its value independent of the
temperature and the magnetic field, provided that the
applied magnetic field is parallel to the mean orientation.
The observed NMR lines are the average over all grains.

We assume that the azimuthal and the polar angles $\theta$
and $\phi$
are given by the Boltzmann distribution,
\begin{equation}
f(\theta,\phi)\propto\exp
\left(-{E_{\rm or}\over k_BT_{\rm prep}}\sin^2\theta\right),
\label{Sl10}
\end{equation}
where $E_{\rm or}$ is the energy needed to disorient the
grain during a preparation at temperature $T_{\rm prep}$.
One can integrate out the polar angle,
\begin{equation}
f(\theta)\propto\sin(\theta) \exp
\left(-{E_{\rm or}\over k_BT_{\rm prep}}\sin^2\theta\right).
\label{Sl11}
\end{equation}
For $E_{\rm or}\gg k_BT_{\rm prep}$, which is a necessary
condition to prepare a well oriented sample, the large angles
are very unlikely and one can use the approximation
\begin{equation}
f(\theta)={2\over \theta_c}\theta\,
{\rm e}^{-{\theta^2/\theta_c^2}},
\label{Sl12}
\end{equation}
where $\theta_c^2=k_BT_{\rm prep}/E_{\rm or}$ measures the
scatter of angles.

The average over the orientation of grains has to be
performed in addition to the natural line width. The
observed distribution of frequencies thus reads,
\begin{equation}
\bar F_{2,3}(\nu)=\int_0^{\pi\over 2} d\theta\, f(\theta)
\int d{\bf r}
{\Gamma\over(\nu-\nu_{2,3}(\theta,{\bf r}))^2+\Gamma^2},
\label{Sl13}
\end{equation}
where $\nu_{2,3}$ is given by (\ref{Sl3}) and (\ref{Sl4}).

The integral over $\theta$ in (\ref{Sl13}) can be evaluated
analytically in terms of the exponential integral of complex
arguments. This step, however, is not favorable numerically.
A faster numerical scheme is obtained if one first evaluate
the distribution $F_{2,3}$ of an ideally oriented sample via
(\ref{NMR3}) with the NMR frequency of a perfectly aligned
magnetic field (\ref{NMR1}). The angular averaging is then
covered by a convolution,
\begin{equation}
\bar F_{2,3}(\nu)=\int_0^{\pi^2\over 4\theta_c^2} dx
\,{\rm e}^{-x}
F_{2,3}\left(\nu\mp{3\over 2}\nu_0\theta_c^2 x\right).
\label{Sl14}
\end{equation} From
the experimental data of KNM we estimate
$\Gamma\approx 140$~kHz and ${3\over 2}\nu_0\theta_c^2=
0.9\Gamma=126$~kHz. For these values the maximum of the
lower NMR frequency is shifted by $78.3$~kHz while the
maximum of the upper frequency is shifted by $-78.3$~kHz.
This shift appears at any temperature and it is subtracted
in the evaluation of the EFG effect. The upper integration
limit, $\pi^2/4\theta_c^2\approx 10^3$, can be replaced by
infinity or a smaller cut off at convenience.

\section{Numerical results}
Our numerical treatment follows the three steps mentioned
above. In the first step we evaluate the GL function
$\psi({\bf r})$ and the magnetic field $B({\bf r})$ of the
Abrikosov vortex lattice using Bardeen's extension of the GL
theory represented by the set of equations
(\ref{C16k0},\ref{SE}-\ref{C14psi}). In this step we take
the magnetic field as parallel to the axis $c$ neglecting
its small perpendicular component in individual grains.

In the second step we use Blatter's approximation (\ref{ephi4})
to evaluate the electrostatic potential in planes from the wave
function. The charge density is obtained from the Poisson
equation (\ref{LS20}) and is rescaled to the transfered number
$N({\bf r})$ of holes per Cu site via (\ref{LS8}).

In the third step we evaluate the local NMR frequencies for
the magnetic field parallel to the axis $c$ using (\ref{Sl4})
with $\theta=0$. The distribution $F$ of the NMR frequencies
for a single crystal with ${\bf B}\parallel c$ is obtained
via space averaging of Lorentzian lines (\ref{NMR3}). The
distribution $\bar F$ of the NMR frequencies for the granular
sample is obtained from $F$ via the convolution (\ref{Sl14}).
The maximum frequencies $\bar\nu_{2,3}$ of the crystal and
granular samples are found from the maxima of $F_{2,3}$ and
$\bar F_{2,3}$, respectively. The latter one can be compared with
the experimental results of KNM, the former represents our
prediction for similar measurements that, as we hope, will
be performed on single crystals in future.

The material parameters of the YBa$_2$Cu$_3$O$_7$ from
Ref.~\onlinecite{P95} are: the density of pairable holes
$n=5\times 10^{27}$~m$^{-3}$, the effective mass $m_p^*=6.92\,m_e$,
the critical temperature $T_c=90$~K, and the Sommerfeld
constant $\gamma_p=302$~JK$^{-2}$m$^{-3}$. These values give
a London penetration depth $\lambda_{\rm Lon}=1.4\times 10^{-7}$~m,
an upper critical field $B_{c2}=96.5$~T, and a GL parameter
at the critical temperature $\kappa_0=55$. In Fig.~2.16
Plakida\cite{P95} shows results of Junod\cite{J90} according
to which the charge transfer $-0.03\,e$ from chains to planes
per Cu site leads to a decrease of the critical temperature by
30~K. This corresponds to $\partial\ln T_c/\partial\ln n_p=-
4.82$. We take the permittivity $\epsilon=4 \epsilon_0$. We
do not discuss the YBa$_2$Cu$_4$O$_8$ for which we could not
find relevant material parameters.

\subsection{Space variation of the NMR frequency}
\begin{figure}   
\psfig{file=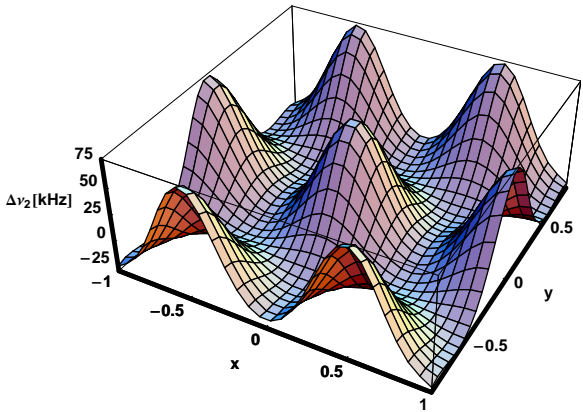,width=8cm}
\caption{The lower NMR frequency $\Delta\nu_2({\bf r})$ in the
crystal at $T=0.6~T_c$. The frequency reflects the triangular
structure of the Abrikosov lattice. The maxima are at the
vortex cores.}
\label{f1}
\end{figure}

Figure~\ref{f1} shows the fishnet plot of the lower NMR
frequency $\Delta\nu_2({\bf r})$ for $T=0.6~T_c$ and $\theta=0$.
The variation is of the order of 100~kHz, which is comparable
to the line width.

\begin{figure}    
\psfig{file=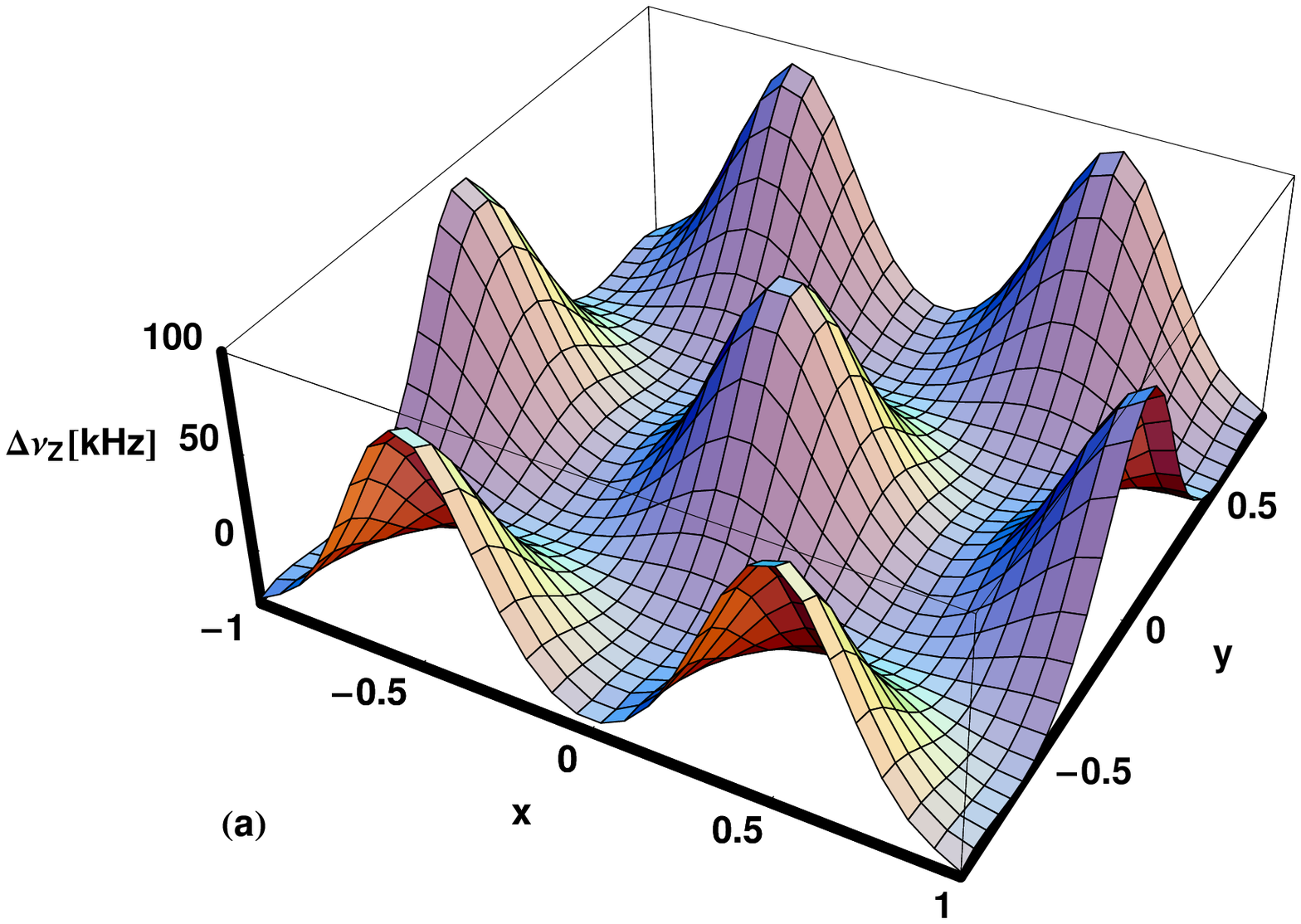,width=8cm}
\psfig{file=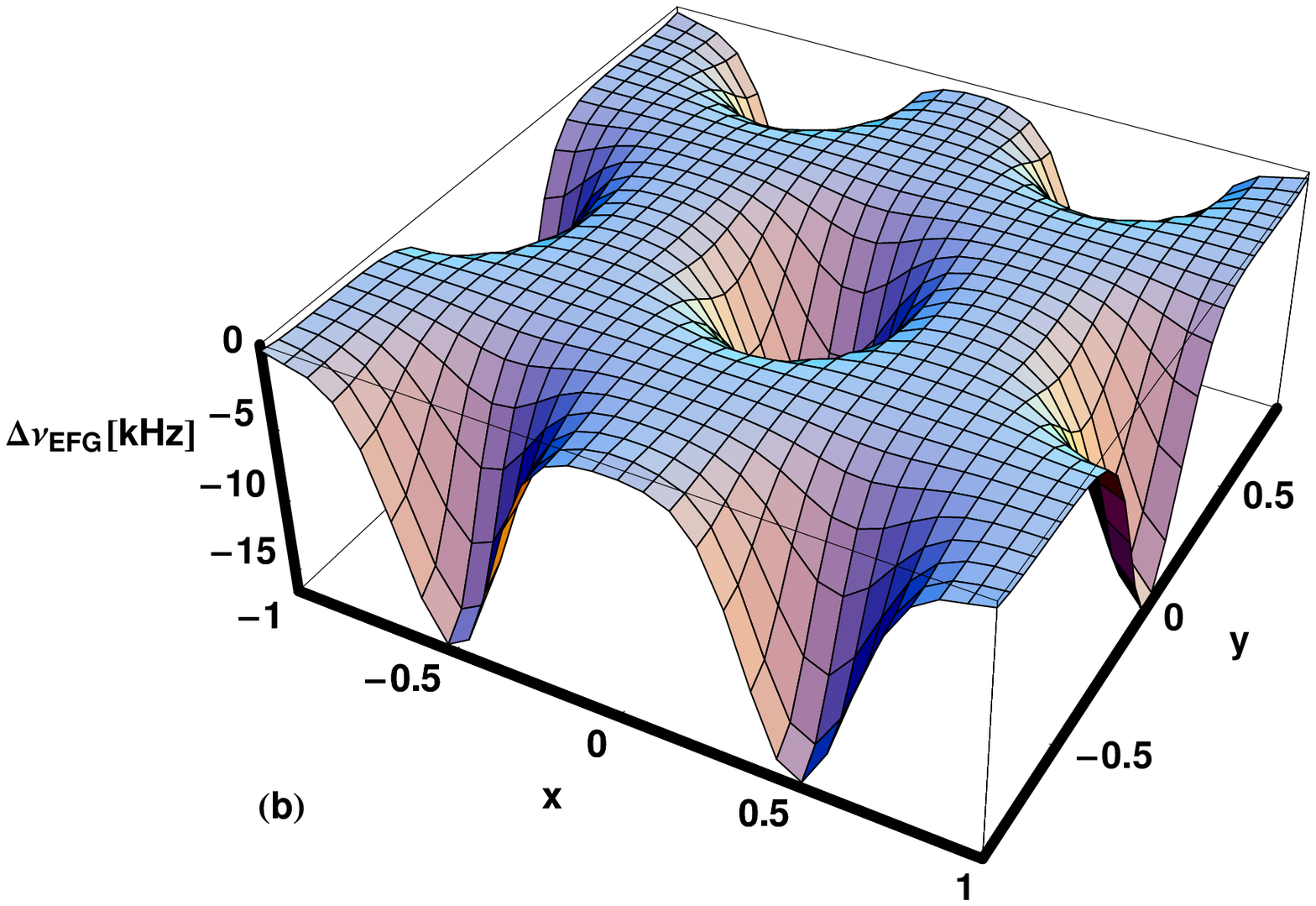,width=8cm}
\caption{The Zeeman effect $\Delta\nu_{\rm Z}({\bf r})=
\gamma_{\rm Cu}\Delta B({\bf r})$ (upper), and the EFG effect
$\Delta\nu_{\rm EFG}({\bf r})=-A\Delta N({\bf r})$ (lower).
Parameters are the same as in Fig.~\protect\ref{f1}.}
\label{f2}
\end{figure}

In Fig.~\ref{f2} we plot the contribution
of the Zeeman effect, $\Delta\nu_{\rm Z}=\gamma_{\rm Cu}
\Delta B$, and the EFG effect, $\Delta\nu_{\rm EFG}=-A\Delta
N$ separately. One can see that the Zeeman effect is about
seven times
larger that the EFG effect. The main contribution to the
spatial variation of the NMR frequency is thus due to the
compression of the magnetic field in vortex cores. On the
other hand, the EFG has nonzero mean values, therefore, it
contributes more to the observable shift of the maxima of
the NMR line.

\begin{figure}  
\psfig{file=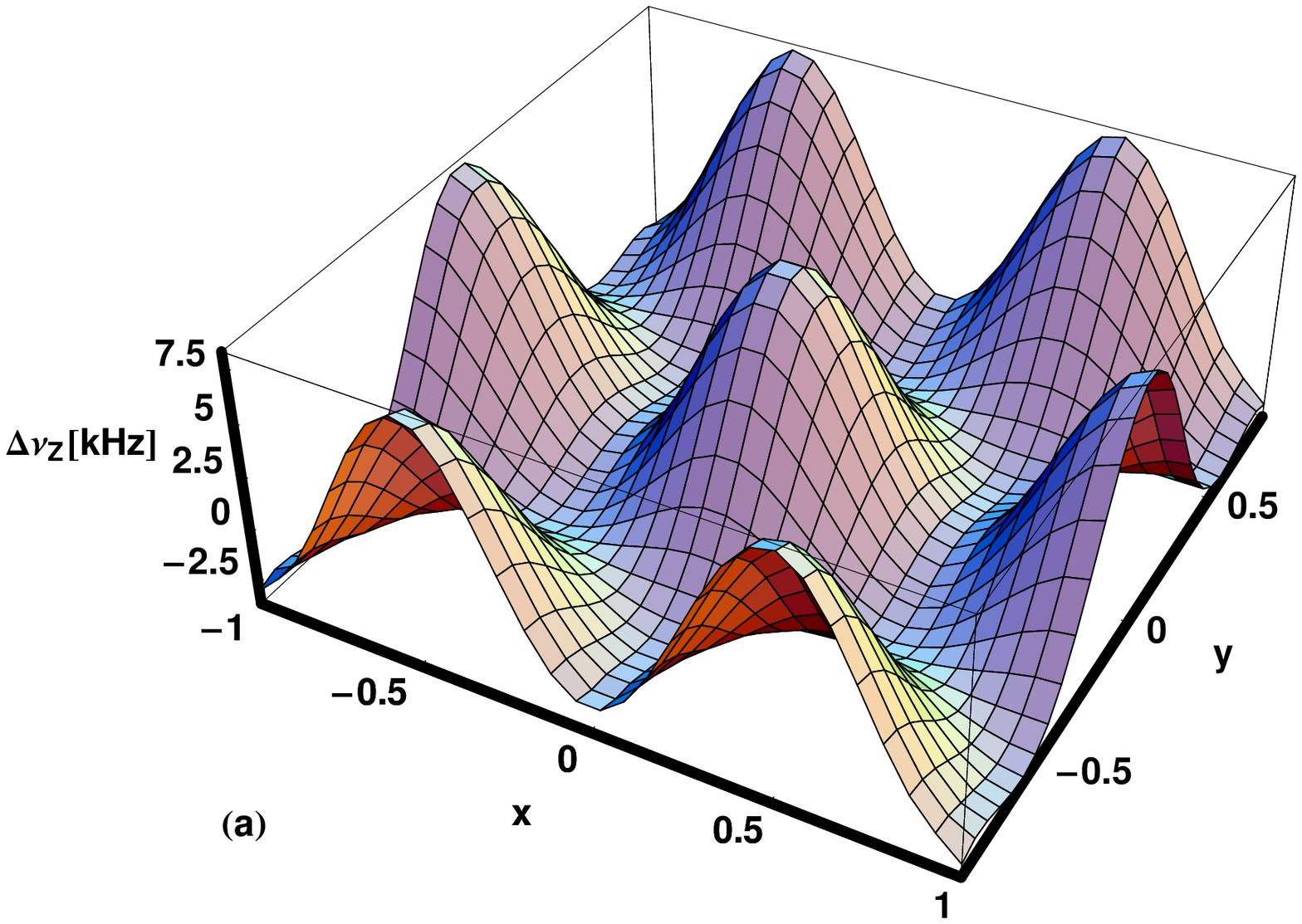,width=8cm}  
\psfig{file=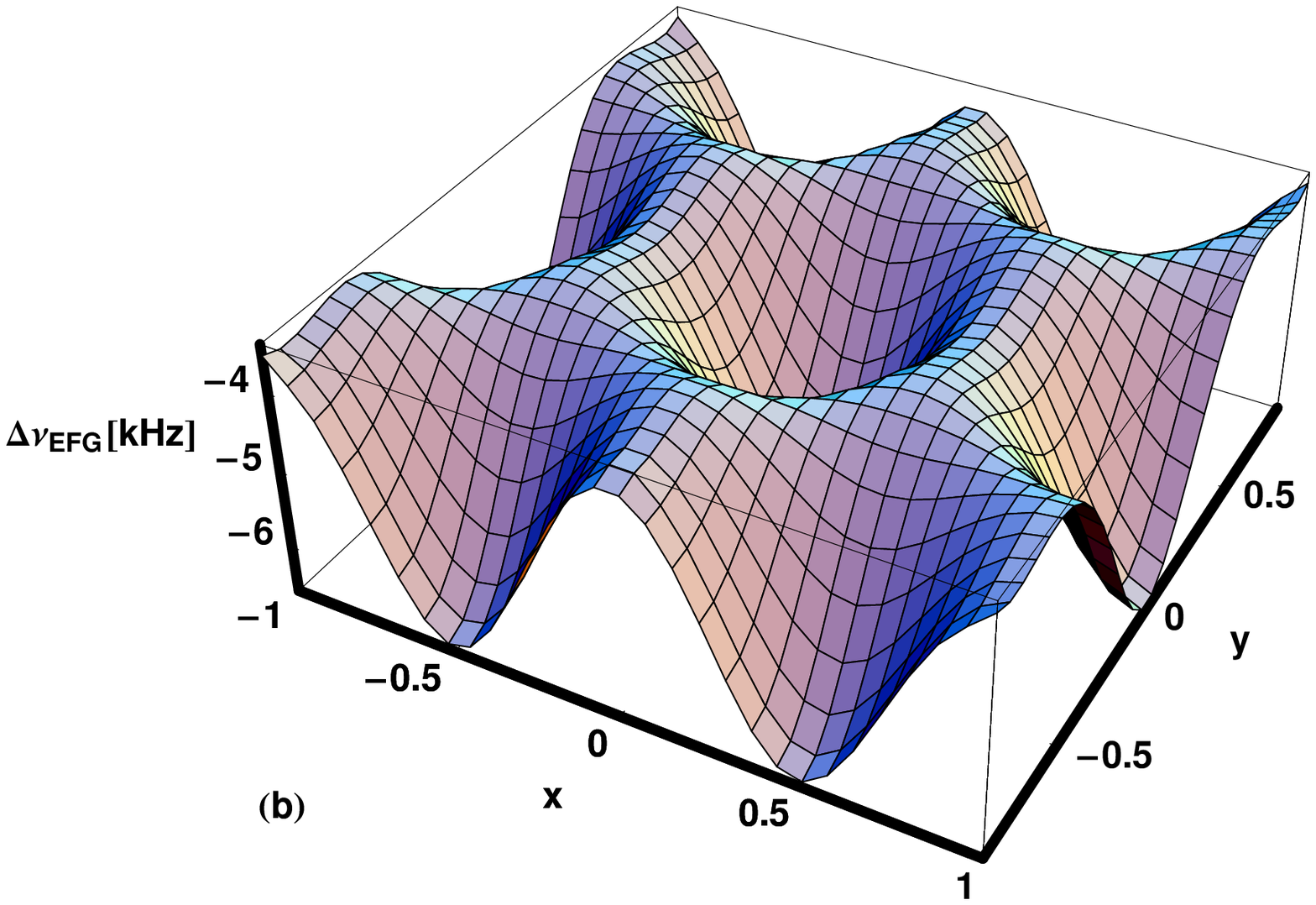,width=8cm}
\caption{The Zeeman effect $\Delta\nu_{\rm Z}({\bf r})=
\gamma_{\rm Cu}\Delta B({\bf r})$ (upper), and the EFG effect
$\Delta\nu_{\rm EFG}({\bf r})=-A\Delta N({\bf r})$ (lower),
for $T=0.9~T_c$.}
\label{f3}
\end{figure}

At temperatures close to $T_c$, the space variation of the
Zeeman effect is reduced following the reduced compression of
the magnetic field in the vortex cores. Fig.~\ref{f3}
shows a decomposition of the lower NMR line into the Zeeman
part and the EFG effect for $T=0.9~T_c$. One can see that
here the amplitude of the EFG effect is close to the amplitude
of the Zeeman effect. Note that the mean value of the EFG is
larger than its spatial variation. This signals that the mean
value provides a good approximation at this temperature.

\subsection{NMR lines of a single crystal}
\begin{figure}  
\psfig{file=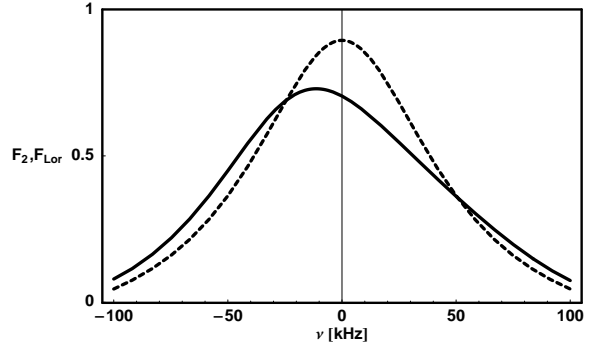,width=8cm}
\caption{The distribution $F_2(\nu)$ (solid line) compared with
the Lorentz distribution $F_{\rm
Lor}(\nu)=\frac 1 \pi {\Gamma\over\nu^2+\Gamma^2}$
(dashed line) at $T=0.6~T_c$ and $\Gamma=50$~kHz. The
experimental line width is $\Gamma=140$~kHz. We use the narrow
line to make its deformation more visible.}
\label{f4}
\end{figure}

Above the critical temperature, the NMR lines of a single
crystal are given by the Lorentzian distribution. Below $T_c$
the space variation of the magnetic field deforms the lines
so that their centers of mass (i.e., mean values) are not
identical to their maxima, see Fig.~\ref{f4}.

The deformation of the NMR line seen in Fig.~\ref{f4} can
be correlated with the space distribution of the NMR frequency
presented in Fig.~\ref{f1}. The vortex cores supply
frequencies in the range from zero to 75~kHz above the mean
value. Their contribution is visible as the extended left
shoulder of the line in Fig.~\ref{f4}. The intermediate region
between vortices supplies frequencies about 25~kHz below the
mean value. Their contribution results in a shift of the
maximum of the NMR line.

\begin{figure}   
\psfig{file=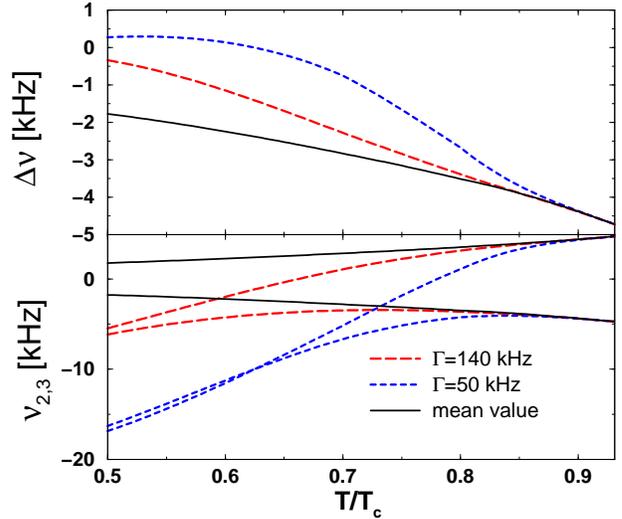,width=8cm,angle=-90}
\caption{The maxima $\bar\nu_{2,3}$ (dashed lines) and mean
values $\langle\Delta\nu_{2,3}\rangle$ (full lines) of the
lower and upper NMR lines (below) as a function of
temperature. The corresponding shifts, $\Delta\bar\nu={1\over 2}
\left(\Delta\bar\nu_2-\Delta\bar\nu_3\right)$ and
$\langle\Delta\nu\rangle={1\over 2}(\langle\Delta\nu_3\rangle-
\langle\Delta\nu_2\rangle)$ are plotted above.}
\label{f5}
\end{figure}

The temperature dependence of the mean values
$\langle\Delta\nu_{2,3}\rangle$ and the
maxima $\Delta\bar\nu_{2,3}$ is presented in Fig.~\ref{f5}.

The mean value is independent of the space variation of the
magnetic field, $\langle\Delta\nu_{2,3}\rangle=
\gamma_{\rm Cu}\langle\Delta B\rangle\mp A\langle\Delta N
\rangle=\mp A\langle\Delta N\rangle$, because $\langle\Delta
B\rangle=0$ by definition. The difference between the mean
value and maximum thus shows the effect of the spatial variation
of the magnetic field on the position of the maximum. Since
the mean value is proportional to the charge transfer while
the position of the maximum is a rather complicated quantity,
the temperature region in the vicinity of $T_c$ is more
recommendable for the experimental exploration. We reserve
the symbol $T_c$ for the critical temperature in the absence
of the magnetic field, $T_c=90$~K. In the magnetic field
$B=9.4$~T, the actual phase transition is at $T=0.931~T_c$.

The deviation of the maximum from the mean value depends on
the line width. This is illustrated in Fig.~\ref{f5} by the
line of artificially chosen width $\Gamma=50$~kHz. In the
limit of very broad lines, $\Gamma\to\infty$, the position
of the maximum becomes identical to the mean value. This
trend is also seen from the formula (\ref{NMR5}).

\subsection{Conventional model}
Let us make a small detour and assume that there is no charge
transfer between chains and planes. In other words, no
screening by chains is assumed and the charge is transfered
exclusively inside the CuO$_2$ planes. In this case the system
would behave as a conventional superconductor in which the
charge is transfered from the vortex cores into the region
between them. In this model, the charge density is given by
$\rho_p=\varphi_p/L_{pp}$, while other relation remain the 
same as in the case of the above layered model.

\begin{figure}    
\psfig{file=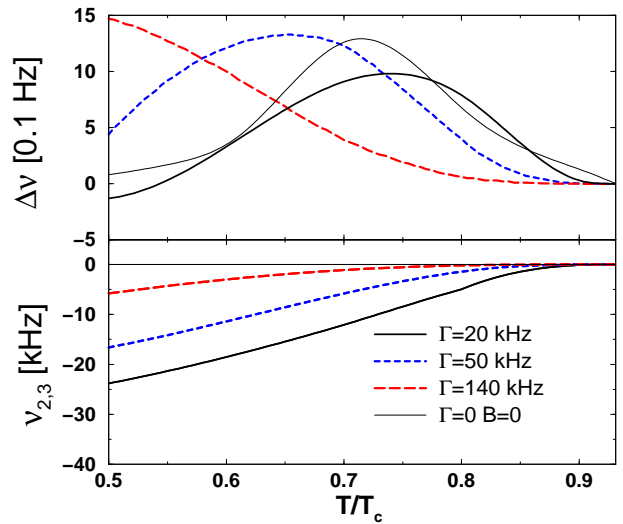,width=8cm,angle=-90}
\caption{The maxima $\bar\nu_{2,3}$ of the lower and upper
NMR lines as a function of temperature (below) and the shift
$\Delta\nu$ (above) in the absence of screening
chains.}
\label{f6}
\end{figure}

One can see in Fig.~\ref{f6} that the sign of $\Delta\nu$ is
opposite to the sign of the model with screening by chains. 
Also, the magnitude of $\Delta\nu$ is extremely small. 
A typical shift is less than 1~Hz while the typical shift 
is more then 1~kHz in the case where the chains are accounted 
for. The values observed by KNM are about 10~kHz. Our result 
thus confirms the estimate of KNM that the conventional model 
of the vortex charge fails to explain their experimental 
data.\cite{KNM01}


When the transfer between chains and planes is prohibited,
the amplitude is reduced by two mechanisms. First, the charge 
transfer accross the vortex core happens on a distance which 
is much larger than the distance between chains and planes. 
Second, the mean value of the shift is zero, $\langle\nu_i
\rangle=0$, because the charge of planes is conserved, 
$N_0-\langle N\rangle=0$. The line shift thus results merely 
from the statistical variations which are small due to the 
large line width, see (\ref{NMR5}). 

An explanation of the reversed sign follows the argument given
already by KNM.\cite{KNM01} Since the magnetic field is much
smaller than $H_{c2}$, the area of vortex cores is much smaller 
than the area between them. Accordingly, cores have a negligible
weight in statistical variations so that the charge between 
cores determines the line shift. This charge is opposite to the 
charge of cores, consequently the reversed sign of the line shift 
results from the conventional model.

\subsection{Averaging over grain orientation}
\begin{figure}     
\psfig{file=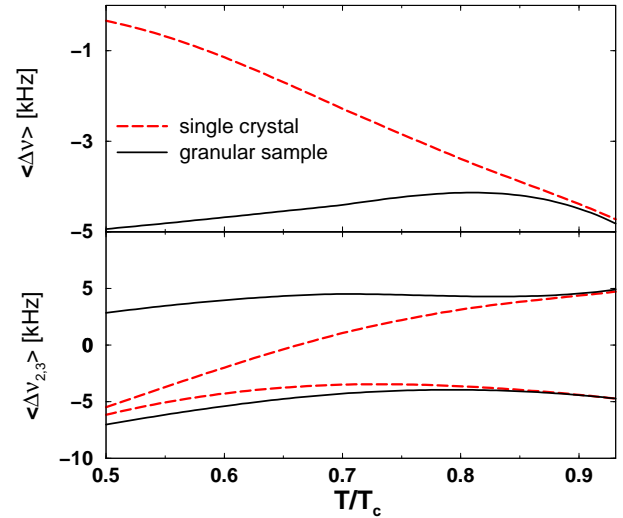,width=8cm,angle=-90}
\caption{The maxima $\Delta\bar\nu_{2,3}$ of the lower and
upper NMR lines as a function of temperature (below) and the
shift $\langle\Delta\nu\rangle$ (above) of $\Gamma=140$ kHz
of Fig.~\protect\ref{f5} compared with the averaging over
grain boundaries according to
(\protect\ref{Sl14}).}
\label{f7}
\end{figure}

As mentioned, KNM have used a granular sample with a high but
not perfect orientation of YBCO grains.\cite{KNM01} The
averaging over the
grain orientation (\ref{Sl14}) results in an additional shift
of the NMR frequencies, see Fig.~\ref{f7}. The grain
disorientation contributes mainly at low temperatures where
it leads to shifts of about 5~kHz, while the value predicted for
the single crystal reduces with temperature.

\begin{figure}    
\psfig{file=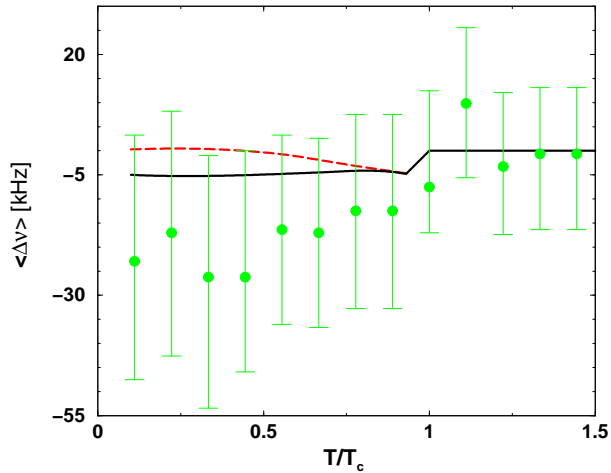,width=8cm,angle=-90}
\caption{Comparison of the theory with experiment.
The experimental data are the dots with error bars, the theoretical
prediction is the full line and the theoretical value for the
single crystal is shown as a dashed line.}
\label{f8}
\end{figure}

Figure~\ref{f8} compares our theoretical results with the
experimental data of KNM. Apparently, the agreement is only
qualitative. We have obtained the correct sign but the
theoretical amplitude is about two or three times smaller
than observed.

In the temperature dependence one can identify three regions.
Above the critical temperature there is no charge
transfer between chains and planes. Indeed, neglecting the
second terms in the potential (\ref{ephi}) we have neglected the
thermally induced charge transfer in the normal state which
appears for unequal energy derivatives of the density of
states, $\partial\gamma_p/\partial n_p\ne \partial\gamma_c/
\partial n_c$. The experimental data of KNM\cite{KNM01} above
the critical temperature justify this neglect.

In the narrow region between $T_c$ and $0.93~T_c$ the sample
is superconducting only if the magnetic field is absent. The
NQR frequency is thus measured in the superconducting state
while the NMR frequencies are measured in the normal state.
In this temperature region the observed frequencies are free
of space fluctuations giving a direct experimental access to
the charge transfer caused by the superconducting transition.
Since in both cases (NMR and NQR) the magnetic field and the
transfered charge are homogeneous, the averaging over the
grain orientation is the same as in the normal state above
$T_c$. Accordingly, the granular sample used by KNM should
provide data comparing well with the single crystal.
Unfortunately, this region is not covered by sufficiently
detailed data in Ref.~\onlinecite{KNM01}.

The region between $0.93~T_c$ and $T_c$ can be used to
experimentally test our assumption about the charge
transfer between chains and planes. In the absence of the
charge transfer between chains and planes, the shift
$\Delta\nu$ is exactly zero since there are no vortices
to cause the in-plane transfer.

Below $0.93~T_c$ the sample is superconducting even in the
presence of the magnetic field, i.e., the Abrikosov vortex
lattice enters the game. Let us first discuss the mean
value. When the magnetic field penetrates the sample, the
superconductivity is suppressed in the vortex cores. The
charge transfer induced by vortices is crudely proportional
to the product of the superconducting fraction and the area
of the vortex core. As the temperature is reduced, the
superconducting fraction grows while the diameter of the
vortex cores shrinks down. For a GL parameter $\kappa$
independent of the temperature, the product remains constant.
Within Bardeen's approximation the GL parameter $\kappa$
slightly decreases with the temperature,\cite{LKMB02}
therefore, the charge transfer reduces as $T\to 0$.

At temperatures below $0.7~T_c$ the maximum of the NMR line
in the granular sample is strongly shifted by the angular
averaging. According to our results, the granular samples in
this temperature region cannot be used to access material
properties.

\section{Summary}
We have shown that the experimental results of Kumagai, Nozaki
and Matsuda can be explained in terms of a standard theory of
the electrostatic field in superconductors extended by
features necessary to describe the layered structure of
YBCO. The agreement of our theory with the experiment is
only qualitative. Our results thus do not exclude
possible alternative mechanisms\cite{S02,CWZT02}.

In conclusion, the charge transfer between CuO chains and
CuO$_2$ planes in YBa$_2$Cu$_3$O$_7$ provides a qualitative
explanation of the electric field gradient effect on the
NMR lines. At low temperatures the effect of the charge
transfer is spoiled by two additional effects: the space
modulation of the Zeeman effect reduces the shift of the
maxima of the NMR lines; the disorientation of grains in
the sample increases the shift. From this point of view it
is advisable to make precise measurement near the critical
temperature where the charge transfer dominates.

Our study does not cover two realistic features that can
increase the amplitude of the observed shift of NMR lines
towards the experimental values. First, the charge is transfered
from the copper to the apex oxygen, i.e., the distance
between planes and chains $D_{cp}=4.24$~\AA\ should be
reduced to the Cu-O distance 2.3~\AA . This will enhance
the charge transfer by a factor 1.8. We did not include
this simple correction as its consistent implementation
requires to use the non-local dielectric function and the
microscopic treatment of electronic states. A similar type
of enhancement would result from a higher value of the
dielectric constant. We have used $\epsilon=4\epsilon_0$
while values as large as $\epsilon=100\epsilon_0$ has 
been advocated.\cite{M92} Second, the
magnetic anisotropy of YBCO enhances the deviations of the
magnetic field from the axes $c$ in individual grains. At low
temperatures this effect should be stronger in agreement
with the observed trends. This correction is omitted because of
an unknown vortex pinning.

It is apparent from the experimental data of KNM that an explanation
of the NMR lines in the underdoped YBa$_2$Cu$_4$O$_8$ will be
more complicated. Unfortunately, we could not discuss
YBa$_2$Cu$_4$O$_8$, because we have not found a relevant set
of material parameters.

\acknowledgments
Authors are grateful to K.~I.~Kumagai, Y.~Matsuda, P.~Nov\'ak
and J.~Rosa for fruitful discussions. This work was supported
by M\v SMT program Kontakt ME601 and Grants Nos. GA\v CR 202000643,
GAAV A1010806 and A1010919. The European ESF program VORTEX
and the support by the Max-Planck society are gratefully acknowledged.

\appendix
\section{Innate perpendicular magnetic field}
The magnetic field perpendicular to the axis $c$ appears in
the sample from two sources. First, the observed samples are
made from many grains with a small but finite scatter of the
orientation of the $c$ axis. This contribution is discussed
in the main text. Second, since diamagnetic currents are in
planes, the magnetic field is more compressed in layers of
planes than in layers of chains. This part of the perpendicular
field is innate to the layered system and appears also in
single crystals.

Here we show that the innate perpendicular magnetic field can be
neglected. The magnetic field is given by the vector potential
${\bf A}$. For ${\bf B}\parallel c$, the component of
${\bf A}$ along the $c$ axis is zero, see (\ref{C16a}).
The perpendicular magnetic field thus reads,
\begin{equation}
B_x=-{\partial A_y\over\partial z},~~~~~~~~~~~~
B_y={\partial A_x\over\partial z}.
\label{Sl5}
\end{equation}
In the short notation, ${\bf B}_\perp=\hat{\bf z}\times
\partial_z {\bf A}$, where $\hat{\bf z}$ is the unit vector in
the direction $z$.

The vector potential (\ref{C16a}) has a cusp at the positions
of layers. This cusp appears due to the singular 2D current
of the selected layer, say at $z_j^+$. From the mirror symmetry
follows that the current in the $z_j^+$-layer does not contribute
to the perpendicular magnetic field in the center of this layer,
i.e., at $z_j^+$ where the Cu nuclei sit. We can thus avoid the
cusp by taking neighbor layers only. The mirror symmetry also
shows that all equivalent $z_{j'}^+$ layers spaced symmetrically
above and below $z_j^+$ give zero net contribution. Accordingly,
the perpendicular magnetic field in the vicinity of $z_0^+$
(for $|z-D_{pp}/2|\ll D_{pp}/2$) is given by the reduced vector
potential,
\begin{equation}
{\bf A}^{\rm red}({\bf r},z)=
{\mu_0\over 4\pi}\int d{\bf r}'
\sum_{j'}{{\bf j}_{j'}^-({\bf r}')
\over\sqrt{|{\bf r'-r}|^2+(z_{j'}^--z)^2}}.
\label{Sl6}
\end{equation}
The 2D Fourier transformation of (\ref{Sl6}) yields
\begin{equation}
{\bf A}^{\rm red}({\bf k},z)={\mu_0\over 2k}
{{\rm e}^{-k(z+D_{pp}/2)}+{\rm e}^{-k(D-z-D_{pp}/2)}
\over 1-{\rm e}^{-kD}}{\bf j}_p({\bf k}).
\label{Sl7}
\end{equation}
The derivative of ${\bf A}^{\rm red}$ with respect to $z$
at $z=D_{pp}/2$ provides us with the perpendicular magnetic
field in terms of the 3D current, ${\bf j}=2{\bf j}_p/D$,
\begin{equation}
{\bf B}_\perp=-{\mu_0D\over 4}
{{\rm e}^{-kD_{pp}}-{\rm e}^{-k(D-D_{pp})}
\over 1-{\rm e}^{-kD}}\hat{\bf z}\times{\bf j}({\bf k}).
\label{Sl8}
\end{equation}

Let us estimate the value of the perpendicular field. To this
end we employ the long wave length limit, $k\to 0$,
\begin{equation}
{\bf B}_\perp\approx -{\mu_0\over 4}(D-2D_{pp})
\hat{\bf z}\times{\bf j}({\bf k}).
\label{Sl9}
\end{equation}
The maximum of the current is at the edge of the vortex core,
i.e., at $r\approx \sqrt{2}\xi$, where $\xi=\lambda/\kappa$
is the GL coherence length. The amplitude of the vector
potential at this point, $A=\Phi_0/(2\pi r)=\Phi_0\kappa/
(2\pi\sqrt{2}\lambda)$, gives the current density
$\mu_0j=A/\lambda^2=\Phi_0\kappa/(2\sqrt{2}\pi\lambda^3)$.
For the YBCO parameters, $(D-2D_{pp})/4=1.33$~\AA, $\kappa=
55$, and $\lambda=1400$~\AA, one finds $B_\perp\approx
6\times\,10^{-4}$~T. This field is far too small shifting the NMR
line by $\nu_0{3\over 2}\sin^2\theta\approx 0.2$~Hz. We have
used $\nu_0=31.5$~MHz.

\end{document}